\documentclass[apl,amsmath,amssymb,amnsfonts,twocolumn,a4paper]{revtex4-1}
\usepackage[T1]{fontenc}
\usepackage{geometry}
\usepackage{graphicx}
\usepackage{color}
\usepackage[usenames,dvipsnames]{pstricks}
\usepackage{epsfig}

\usepackage{setspace}
\usepackage{pgfplots}
\usepackage{booktabs}
\usepackage{tikz, tikz-3dplot}
\tikzset{>=latex}
\usepackage[percent]{overpic}
\usetikzlibrary{pgfplots.groupplots,arrows,decorations.markings}
\usetikzlibrary{shapes.geometric,calc}
\usepackage[strict]{changepage}

\begin{document}





\title{Heat assisted magnetic recording of bit patterned media beyond 10\,Tb/in$^2$} 








\author{Christoph Vogler}
\email{christoph.vogler@tuwien.ac.at}
\affiliation{Institute of Solid State Physics, TU Wien, Wiedner Hauptstrasse 8-10, 1040 Vienna, Austria}
\affiliation{Institute of Analysis and Scientific Computing, TU Wien, Wiedner Hauptstrasse 8-10, 1040 Vienna, Austria}

\author{Claas Abert}
\author{Florian Bruckner}
\author{Dieter Suess}
\affiliation{Christian Doppler Laboratory for Advanced Magnetic Sensing and Materials, Institute for Solid State Physics, TU Wien, Wiedner Hauptstrasse 8-10, 1040 Vienna, Austria}

\author{Dirk Praetorius}
\affiliation{Institute of Analysis and Scientific Computing, TU Wien, Wiedner Hauptstrasse 8-10, 1040 Vienna, Austria}









\date{\today}

\begin{abstract}
The limits of the areal storage density as can be achieved with heat assisted magnetic recording (HAMR) are still an open issue. 
We want to address this central question and present the design of a possible bit patterned medium with an areal storage density above 10\,Tb/in$^2$. The model uses hard magnetic recording grains with 5\,nm diameter and 10\,nm height. It assumes a realistic distribution of the Curie temperature of the underlying material as well as a realistic distribution of the grain size and the grain position. In order to compute the areal density we analyze the detailed switching behavior of a recording bit under different external conditions, which allows to compute the bit error rate of a recording process (shingled and conventional) for different grain spacings and write head positions. Hence, we are able to optimize the areal density of the presented medium.
\end{abstract}

\keywords{heat assisted recording, Landau-Lifshitz-Bloch, graded Curie temperature, shingled writing, bit patterned media, bit error rate}
\maketitle 


\section{Introduction}
\label{sec:intro}
Although the basic concept of heat assisted recording (HAMR) reaches back almost 60 years \cite{mayer_curiepoint_1958} only very recently the first fully functional drive was realized with more than 1000 write power on hours \cite{rausch_hamr_2013,rausch_recording_2015} and a 1.4\,Tb/in$^2$ device was demonstrated \cite{ju_high_2015}. In order to keep up the continuous increase of areal storage density (AD) the following factors are essential (i) provide small magnetic grains and (ii) provide a recording scheme with a high effective write field and temperature gradient in order to allow for small bit transitions. To realize small magnetic grains high magnetic anisotropy has to be used to ensure that the stored binary information is thermally stable. The limited maximum magnetic field of write heads results in the so called magnetic recording trilemma. HAMR \cite{mee_proposed_1967,guisinger_thermomagnetic_1971,kobayashi_thermomagnetic_1984,rottmayer_heat-assisted_2006} can help to overcome this trilemma. One uses a laser spot to locally heat the selected recording bit near or above the Curie temperature ($T_{\mathrm{C}}$). Hence, even the magnetization of very hard magnetic materials can be reversed with the available write fields. Nevertheless, thermally written-in errors are a serious problem of HAMR \cite{richter_thermodynamic_2012}. In this paper we will show under which circumstances a bit patterned recording medium, consisting of hard magnetic single phase grains, can have an areal storage density of 10\,Tb/in$^2$ and more, despite thermal fluctuations at high temperatures during writing, which significantly deteriorates the bit transition, as well as the distribution of the Curie temperature of the recording grain are considered. 

\section{Model}
\label{sec:models}
The accurate calculation of the magnetic behavior of a recording grain is a challenging task. During HAMR temperatures near and above $T_{\mathrm{C}}$ of the involved materials can occur, and thus a meaningful physical model should be able to reproduce the phase transition from a ferromagnetic to a paramagnetic state at $T_{\mathrm{C}}$. We use the Landau-Lifshitz-Bloch (LLB) equation for this purpose, which has already been validated in different publications~\cite{garanin_thermal_2004,chubykalo-fesenko_dynamic_2006,atxitia_micromagnetic_2007,kazantseva_towards_2008,chubykalo-fesenko_dynamic_2006,schieback_temperature_2009,bunce_laser-induced_2010,evans_stochastic_2012,mcdaniel_application_2012,greaves_magnetization_2012,mendil_speed_2013}. In this work we use the formulation proposed by Evans~et~al.~\cite{evans_stochastic_2012}. To accelerate the simulations in order to provide an insight into the detailed switching behavior of realistic recording grains we use a coarse grained approach. For detailed information about the used coarse grained LLB model please refer to \cite{volger_llb}. In summary, in the coarse grained LLB model each material is described with just one magnetization vector. With this approach recording grains with realistic lateral dimensions of several nanometers can be efficiently simulated with low computational effort. Nevertheless, the resulting dynamic trajectories reproduce the according computationally very expensive atomistic simulations.

\begin{table}[h!]
  \centering
  \begin{tabular}{c c c c}
    \toprule
    \toprule
      $K_1$\,[J/m$^3$] & $J_{\mathrm{S}}$\,[T] & $A_{\mathrm{ex}}$\,[pJ/m] & $T_{\mathrm{C}}$\,[K] \\
    \midrule
      $6.6\cdot10^6$ & 1.43 & 21.58 & 536.94\\
    \bottomrule
    \bottomrule
  \end{tabular}
  \caption{\small Magnetic properties of the HM recording grain.}
  \label{tab:prop}
\end{table}
In this work we investigate recording grains with a diameter of 5\,nm and a height of 10\,nm. They consist of a hard magnetic (HM) material with uniaxial anisotropy and strong exchange coupling. The detailed material parameters can be found in Tab.~\ref{tab:prop}.

The heat source is assumed to be a continuous Gaussian shaped heat pulse in space and time, which moves over the recording medium, if not stated differently, with a velocity of $v=7.5$\,m/s. The pulse is modeled with
\begin{equation}
\label{eq:gauss_profile}
 T(x,y,t)=\left ( T_{\mathrm{write}}-T_{\mathrm{min}} \right )e^{-\frac{\left (x-vt   \right )^2+\left (y-y_0   \right )^2}{2\sigma^2}}+T_{\mathrm{min}}
\end{equation}
being
\begin{equation}
 \sigma=\frac{\mathrm{FWHM}}{\sqrt{8\ln(2)}}\text{,}\quad T_{\mathrm{peak}}=\left ( T_{\mathrm{write}}-T_{\mathrm{min}} \right )e^{-\frac{\left (y-y_0   \right )^2}{2\sigma^2}}+T_{\mathrm{min}}\nonumber
\end{equation}
where FWHM is the full width at half maximum of the spatial Gaussian, $x$ is the down-track position of a bit and $y-y_0$ is the off-track distance between heat spot and bit. Depending on the off-track distance the temporal heat pulse reaches different peak temperatures $T_{\mathrm{peak}}$. State of the art near field transducers allow a narrow FWHM of 20\,nm. All performed simulations start with an initial temperature of $T_{\mathrm{min}}=270$\,K. The external magnetic field aligning the magnetization of a bit is assumed to be homogeneous in space with a strength of 0.8\,T and is modeled with a trapezoidal shaped pulse with a duration of 1\,ns and a field rise and decay time of 0.1\,ns, respectively. At the start and the end of the simulation time the field points down and in the middle it switches to the up direction, yielding a symmetric pulse in time. The down-track position $x$ as well as the off-track distance $y-y_0$, and thus $T_{\mathrm{peak}}$  are of great importance for a successful write process with a small bit error rate. 
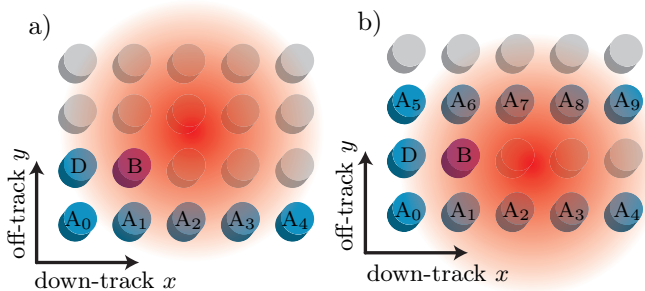
\begin{figure}[!h]
\begin{minipage}{\linewidth}
\vspace{0.05\linewidth}
\begin{minipage}{0.45\linewidth}
  \vspace{-0.22\linewidth}
  \begin{overpic}[scale=.25]{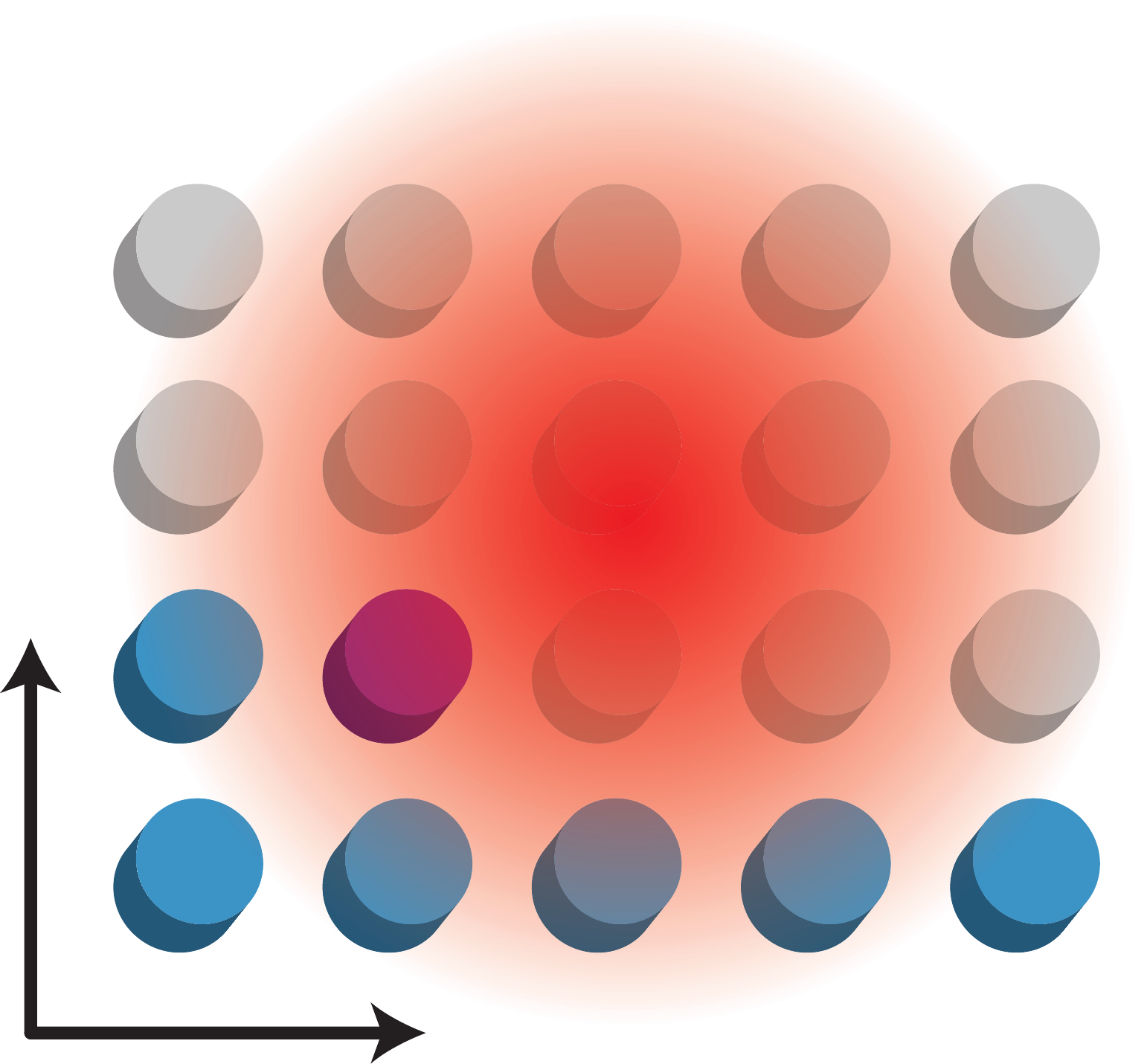}
  \put(0,80){a)}
  \put(2,-6){\small down-track $x$}
  \put(-7,4){\small \rotatebox{90}{off-track $y$}}
  \put(33,33){\footnotesize B}
  \put(14,33){\footnotesize D}
  \put(12,15){\footnotesize A$_0$}
  \put(31,15){\footnotesize A$_1$}
  \put(49,15){\footnotesize A$_2$}
  \put(67,15){\footnotesize A$_3$}
  \put(85,15){\footnotesize A$_4$}
  \end{overpic}
\end{minipage}
\hspace{0.05\linewidth}
\begin{minipage}{0.45\linewidth}
  \begin{overpic}[scale=.25]{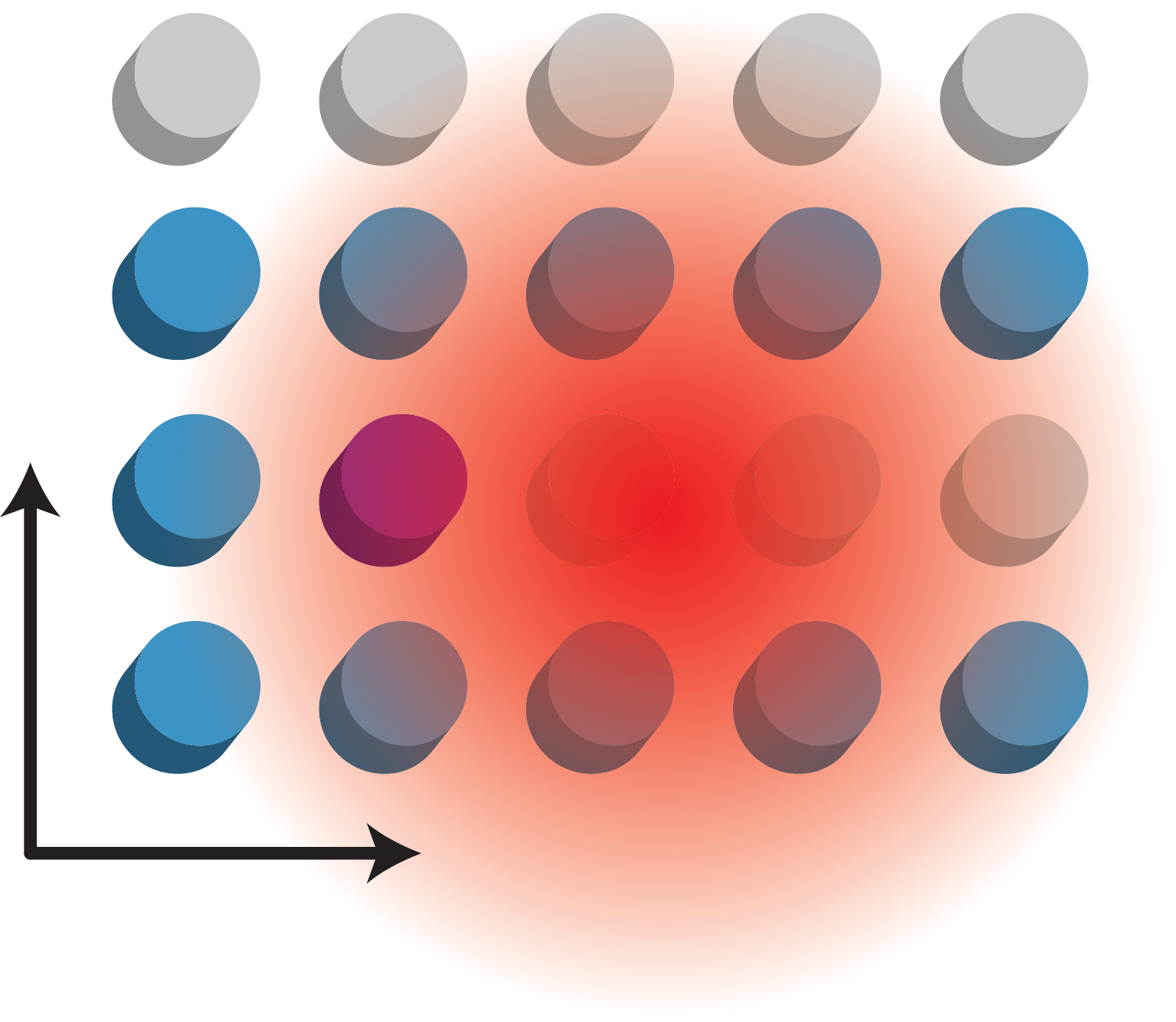}
  \put(0,87){b)}
  \put(3,5){\small down-track $x$}
  \put(-7,15){\small \rotatebox{90}{off-track $y$}}
  \put(32,44){\footnotesize B}
  \put(14,44){\footnotesize D}
  \put(12,26){\footnotesize A$_0$}
  \put(30,26){\footnotesize A$_1$}
  \put(47,26){\footnotesize A$_2$}
  \put(65,26){\footnotesize A$_3$}
  \put(83,26){\footnotesize A$_4$}
  \put(12,62){\footnotesize A$_5$}
  \put(30,62){\footnotesize A$_6$}
  \put(47,62){\footnotesize A$_7$}
  \put(65,62){\footnotesize A$_8$}
  \put(83,62){\footnotesize A$_9$}
  \end{overpic}
\end{minipage}
\end{minipage}
    \caption{\small (color online) Schematic illustration of the bits involved during a) shingled and b) conventional HAMR. B denotes the bit which has to be written. The down-track bit D and all adjacent bits A$_i$ have to remain in their original state.}
  \label{fig:writer_pattern}
\end{figure}
As illustrated in Fig.~\ref{fig:writer_pattern} we examine the areal density of both conventional as well as shingled recording. The recording head, including the heat spot with a temperature of $T_{\mathrm{write}}$, moves in down-track direction over the bits and tries to write bit B. Along one track the peak temperature $T_{\mathrm{peak}}$ is the same for all grains.

\section{Results}
\label{sec:results}
\begin{figure}[!h]
\includegraphics[width=1\linewidth]{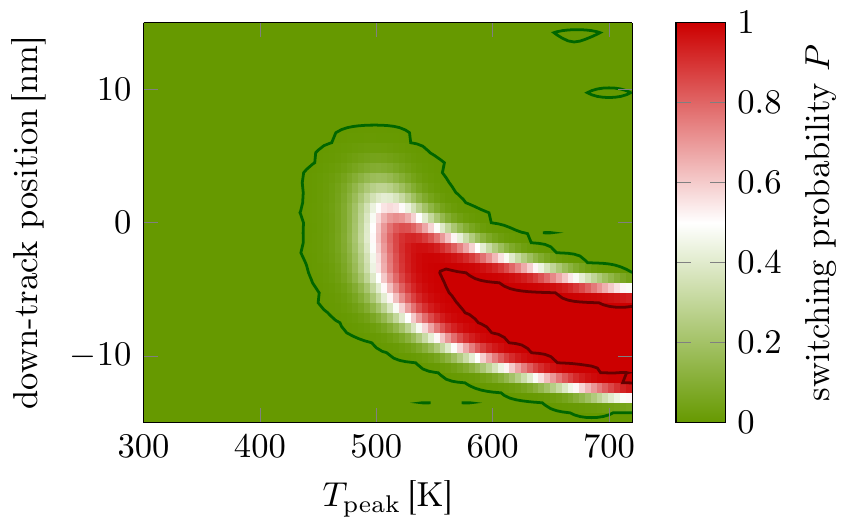}
  \caption{\small (color online) Switching probability phase diagram for a single phase HM grain. An external magnetic field of 0.8\,T is applied to the grains. Each phase point consists of 128 switching trajectories with the same down-track position and the same peak temperature. The solid lines around the phase transitions separate the dark red areas, where complete switching occurs ($P>99.9$\,\%) and the light green areas, where no switching is possible ($P<0.1$\,\%), or where the grain switches in $-z$ and back to $+z$-direction, which finally also counts as non-switched. The phase diagram already contains a distribution of the Curie temperature of $\sigma_{T_{\mathrm{C}}}=3$\,\%\,$T_{\mathrm{C}}$.}
  \label{fig:clsr_phase}
\end{figure}
Our goal is to investigate the switching behavior of the HM grain in order to calculate the areal density of a bit patterned medium. Since the coarse grained LLB approach is fast and reliable, a detailed phase diagram of the switching probability $P$ can be simulated, as illustrated in Fig~\ref{fig:clsr_phase}. Here the influence of the peak temperature of the heat source and its down-track position is examined. In each simulation a full heat pulse is applied to the grain with an additional external field pulse, trying to switch the particle from the $+z$ to the $-z$-direction. After the simulation the state of the particle, if it has switched or not, is evaluated. In each phase point 128 trajectories are calculated to obtain a switching probability. The resolution of the diagram in the temperature axis is 5\,K and in the down-track position axis 0.75\,nm. Hence, the diagram contains the data of about 450 000 switching trajectories. The contour lines near the phase transitions separate the dark red areas with $P>99.9$\,\% from the light green areas with $P<0.1$\,\%. A distribution of the Curie temperature of $\sigma_{T_{\mathrm{C}}}=3$\,\%\,$T_{\mathrm{C}}$ is considered. The diagram shows a core with complete switching above the Curie temperature and at a down-track position of about $-7.5$\,nm. The obtained plot can be interpreted as the footprint of the recording head. Along the y-axis in the red area all grains are written in the $-z$ direction and in the green area the grains are pointing in the $+z$-direction (either because they were written from $+z$ to $-z$ and back to $+z$, or they could not be reversed at all and are still in the $+z$-direction). 

Even more important than the switching distribution is the knowledge of the bit error rate (BER) during a recording process. With the data of the detailed switching phase diagram in Fig.~\ref{fig:clsr_phase} it is possible to directly calculate the BER for shingled and conventional HAMR for a bit patterned medium. Figure~\ref{fig:writer_pattern} compares the relevant bits for the calculation of the BER according to 
\begin{equation}
\label{eq:BER}
 \mathrm{BER}=1-P_{\mathrm{B}}\left ( 1-P_{\mathrm{D}} \right )\prod_{i=0}^{j}\left ( 1-P_{\mathrm{A_i}} \right )^{n}.
\end{equation}
The total BER of bit B contains the product of the joint probability to successfully write bit B, but not the previous bit D on-track and not the adjacent bits A$_i$ down-track, which have to remain in the same magnetic state. The exponent $n$ in Eq.~\ref{eq:BER} denotes the number of write processes for which all adjacent bits have to retain their state. During shingled recording the information of a whole block is written at once. Hence, just the adjacent bits A$_0$ to A$_4$ in Fig.~\ref{fig:writer_pattern}a) have to be considered in Eq.~\ref{eq:BER} with $n=1$. During the conventional recording technique one track is written. Thus, one has to account for the adjacent bits on both sides off-track to compute the BER. Furthermore, it is required that the adjacent bits are stable for at least $n=1000$ write processes. To calculate the switching probability $P$ of each bit the relative down-track position of the bit and the heat spot and their off-track distance, in order to determine the peak temperature of the heat pulse at the track, have to be computed. The correct switching probability can then be obtained from Fig.~\ref{fig:clsr_phase}. As a consequence the BER is calculated for an arbitrary heat spot position according to Eq.~\ref{eq:BER}. We also account for a grain size and displacement distribution ($\sigma_{\mathrm{bitSize}}$ and $\sigma_{\mathrm{bitPos}}$) as well as a distribution of the write head position ($\sigma_{\mathrm{headPos}}$) in agreement with the guidelines of the Advanced Storage Technology Consortium (ASTC)
\begin{eqnarray}
\label{eq:displacement_jitter}
  &\sigma_{\mathrm{bitSize}}&=5\,\%\,\min(l_x,l_y)\nonumber \\
  &\sigma_{\mathrm{bitPos}}&=5\,\%\,\min(l_x,l_y)\nonumber \\
  &\sigma_{\mathrm{headPos}}&=2\,\%\,l_x,
\end{eqnarray}
where $l_x$ and $l_y$ are the center to center spacings between neighboring recording bits in $x$- and $y$-direction, respectively. Hence, the total displacement jitter is
\begin{equation}
  \label{eq:displacement_jitter_tot}
  \sigma_{\mathrm{displ}}=\sqrt{\sigma_{\mathrm{bitSize}}^2+\sigma_{\mathrm{bitPos}}^2+\sigma_{\mathrm{headPos}}^2}.
\end{equation}
In our analysis of the BER we can optimize two parameters in order to maximize the areal storage density AD:
\begin{itemize}
 \item the write temperature $T_{\mathrm{write}}$ of the heat source for a given spacing of the recording grains and 
 \item the the off-track spacing $l_y$ of the medium for a given write temperature.
\end{itemize}
\begin{figure}
\includegraphics[width=1\linewidth]{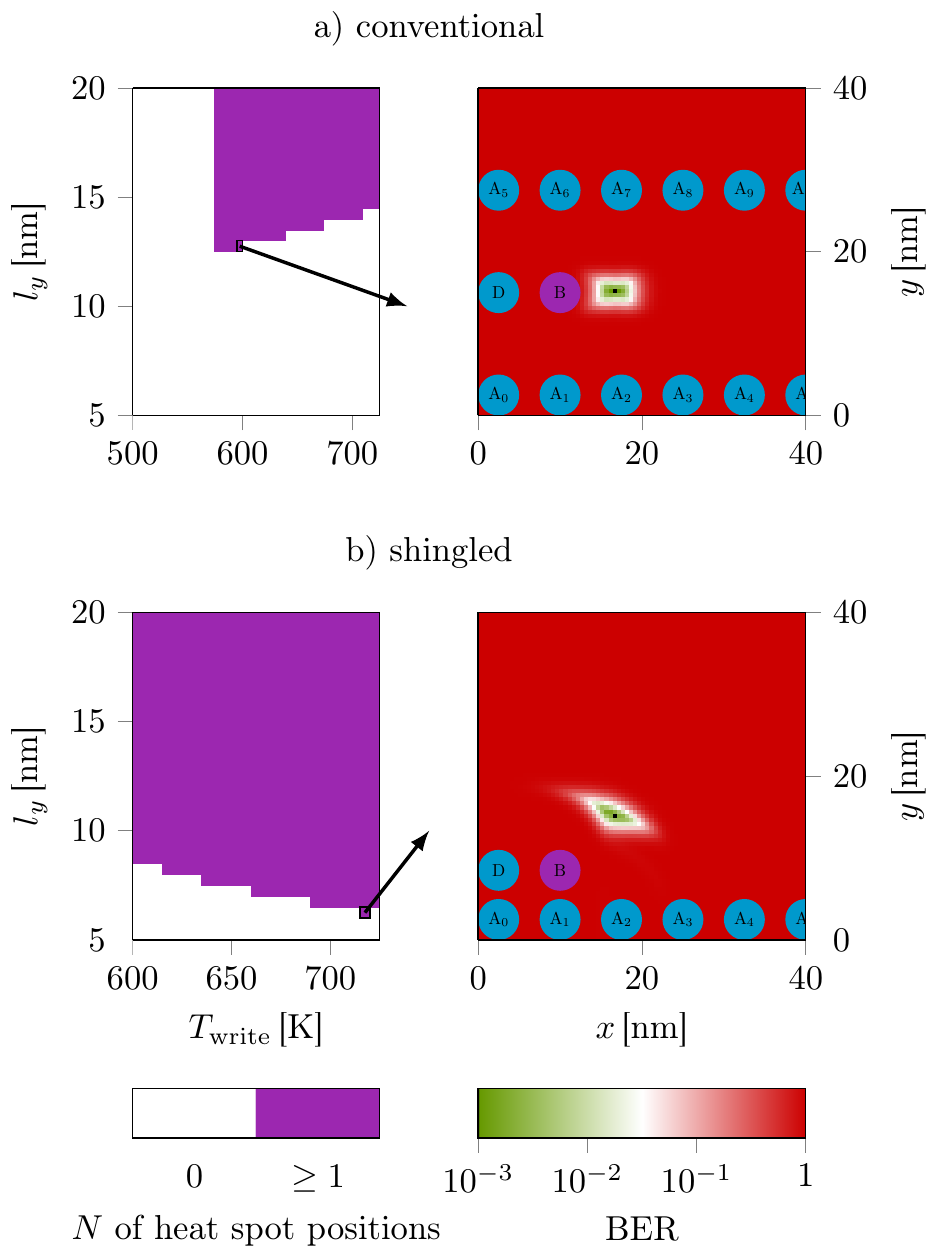}
  \caption{\small (color online) Areal density optimizations of a) shingled and b) conventional HAMR of a HM grain at a write field of 0.8\,T. On the left a map of possible off-track grain spacings and write temperatures to successfully switch bit B under the constraint of $\text{BER}<10^{-3}$ is shown. The white area illustrates parameters where $\text{BER}\ge10^{-3}$ for all heat spot positions and in the purple area at least one position with $\text{BER}<10^{-3}$ can be found. A displacement jitter according to Eqs.~\ref{eq:displacement_jitter} and \ref{eq:displacement_jitter_tot} is considered. In case of the grain arrangement with the highest possible areal density, the BER for different heat spot positions $x$ and $y$ is shown in detail on the right. The black points display positions where $\text{BER}<10^{-3}$. The bit B has to be written while all neighboring bits D and A$_i$ should retain their state.}
  \label{fig:clsr_BER}
\end{figure}
The down-track distance is determined by $l_x = vt_{\mathrm{field}}=7.5$\,nm, where $t_{\mathrm{field}}=1$\,ns is the duration of the magnetic field pulse. The left plots of Fig.~\ref{fig:clsr_BER} show the optimizations for both shingled and conventional recording. These plots distinguish between pairs of $T_{\mathrm{write}}$ and $l_y$ for which at least one heat spot position can be found in order to guarantee $\mathrm{BER}<10^{-3}$ and for which $\mathrm{BER}\ge 10^{-3}$ for all positions. Under the requirement of displacement and size distributions according to Eq.~\ref{eq:displacement_jitter_tot} it is just required to find a single heat spot position with $\mathrm{BER}<10^{-3}$. For conventional recording $l_y$ increases for decreasing write temperatures as long as $T_{\mathrm{write}}$ is large enough to write bit B with a high probability. This fact is not surprising because for a fixed down-track spacing the adjacent tracks can be placed closer to the write track for smaller heat spot temperatures without any risk of adjacent track erasures. We see a contrary behavior for shingled recording. Here the heat spot can be placed off-track and one can use the higher temperature gradient of the heat pulse between the tracks to significantly reduce the off-track distance. 

In case of maximum AD the plots at the right-hand-side of Fig.~\ref{fig:clsr_BER} display the map of the BER for different positions of the heat source. As an illustration also the involved bits are shown. The black points indicate the head positions with $\mathrm{BER}<10^{-3}$. The BER maps confirm the above statements.
\begin{table}[h!]
  \centering
  \vspace{0.5cm}
  \begin{tabular}{c c c c c | c}
    \toprule
    \toprule
      & $l_x$\,[nm] & $l_y$\,[nm] & $T_{\mathrm{write}}$\,[nm] & $\sigma_{T_{\mathrm{C}}}$\,[\%\,$T_{\mathrm{C}}$] & AD\,[Tb/in$^2$]  \\
    \hline
      SR & 7.5 & 6.0 & 715.0 & 3 & 14.34\\
      CR & 7.5 & 12.5 & 595.0 & 3 & 6.88\\
    \hline
      SR & 10.0 & 6.5 & 720.0 & 3 & 9.93\\
      CR & 10.0 & 13.0 & 595.0 & 3 & 4.96\\
    \hline
      SR & 7.5 & 6.5 & 715.0 & 3.15 & 13.23\\
      CR & 7.5 & 13.0 & 595.0 & 3.15 & 6.62\\    
    \bottomrule
    \bottomrule
  \end{tabular}
  \caption{\small Optimal areal storage densities and the corresponding off-track spacings and write temperatures for shingled (SR) and conventional (CR) bit patterned recording considering a $T_{\mathrm{C}}$ distribution of the recording bits.}
  \label{tab:clsr_BER}
\end{table}
Table~\ref{tab:clsr_BER} shows the final maximum AD with the corresponding parameters. Shingled recording reaches the remarkable density of 14.34\,Tb/in$^2$, which is more than twice the AD than for conventional recording. Additionally we simulated the same footprint as in Fig.~\ref{fig:clsr_phase} for a head velocity of 10\,m/s fixing $l_x$ to 10\,nm. The optimized AD for these calculations can also be found in Tab.~\ref{tab:clsr_BER} for comparison.

So far no interactions between the recording grains were considered, as the switching probabilities of Fig.~\ref{fig:clsr_phase} are based on trajectories of just one bit. Due to the small distances between the bits, the demagnetization field of neighboring grains can influence the effective write field of the recording head. Our simulations show that a change in the applied magnetic field shifts the phase transition in Fig.~\ref{fig:clsr_phase} along the peak temperature axis. Additionally a slight broadening of the transition area for larger fields can be mentioned. The same effect is caused by a change of the Curie temperature of the material. Hence, we are able to include the interaction of the recording bits due to their demagnetization field in an additional distribution of $T_{\mathrm{C}}$. To quantify this effect we inspect a complete finite element model of bit B and its 24 nearest neighbors with $l_y=7.5$\,nm and $l_y=6$\,nm. The total field at B is calculated for 50000 configurations consisting of randomly chosen magnetization down and up directions of the neighbors. A histogram of the resulting field values allows to extract the underlying distribution and its standard deviation. With the computed shifts of the transitions in the switching probability diagrams for different applied fields the field distribution can be transferred to an additional distribution of the Curie temperature. In total an increase of 5\,\% in $\sigma_{T_{\mathrm{C}}}$ is obtained. As a consequence the AD of the bit patterned medium slightly decreases to 13.23\,Tb/in$^2$ for shingled and 6.62\,Tb/in$^2$ for conventional recording as shown in Tab.~\ref{tab:clsr_BER}. Even in the more conservative scenario of a 10\,\% increase of the $T_{\mathrm{C}}$ distribution to $\sigma_{T_{\mathrm{C}}}=3.3$\,\%\,$T_{\mathrm{C}}$ the final AD does not significantly change. 
%

\section{Conclusion}
\label{sec:conclusion}
In summary, we used an efficient coarse grained LLB model \cite{volger_llb} to calculate switching probabilities and bit error rates of hard magnetic (FePt like) recording bits with a diameter of 5\,nm and a height of 10\,nm during HAMR. In our calculations we considered a distribution of the Curie temperature of the recording grains, as well as a distribution of the grain size, its position and the position of the heat spot on the medium. We also included interactions between the bits in our calculations as additional contribution to the intrinsic $T_{\mathrm{C}}$ distribution. Hence, we obtained a realistic model of HAMR, where we optimized the areal density in terms of the spacings between the bits and the write temperature of the heat spot. Based on this model we presented a bit patterned recording medium, which reaches an areal storage density of 13.23\,Tb/in$^2$ for shingled and 6.62\,Tb/in$^2$ for conventional recording, if a heat spot with a full width at half maximum of 20\,nm and a velocity of 7.5\,m/s and an external magnetic field of 0.8\,T is assumed.

\section{Acknowledgements}
The authors would like to thank the Vienna Science and Technology Fund (WWTF) under grant MA14-044, the Austrian Science Fund (FWF): F4102 SFB ViCoM and the Advanced Storage
Technology Consortium (ASTC) for financial support. The support from the CD-laboratory AMSEN (financed by the Austrian Federal Ministry of Economy, Family and Youth, the National Foundation for Research, Technology and Development) was acknowledged. The computational results presented have been achieved using the Vienna Scientific Cluster (VSC).


%

\end{document}